\pdfoutput=1
\documentclass[pdflatex,sn-mathphys-num]{sn-jnl}

\usepackage[utf8]{inputenc}
\usepackage{fancyhdr}
\usepackage[fleqn]{amsmath}
\usepackage{amssymb}
\usepackage{amsthm}
\usepackage{bm}
\usepackage{colonequals}
\usepackage{overpic}
\usepackage{url}
\usepackage{xspace}

\usepackage{environ}
\usepackage{enumitem}
\usepackage{listings}
\usepackage{float}
\usepackage{minibox}
\usepackage{fnpct}

\renewcommand{\orcidlogo}{\includegraphics[height=10pt]{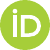}}
\renewcommand{\orcid}[1]{\href{https://orcid.org/#1}{\orcidlogo}}

\usepackage{tikz}
\usetikzlibrary{arrows}
\tikzset{
    treenode/.style = {
        align=center,
        inner sep=0pt,
        text centered,
        font=\sffamily,
        rectangle,
        rounded corners=3mm,
        draw=black,
        minimum width=2em,
        minimum height=2em,
        inner sep=1mm,
        outer sep=0mm
    },
    smalltext/.style = {
        font=\footnotesize
    },
    smalltreenode/.style = {
        treenode, smalltext
    },
    nonuniformtreenode/.style = {
        smalltreenode,
        rectangle,
        rounded corners=0mm
    },
    uniformtreenode/.style = {
        smalltreenode,
        circle
    },
    basisnode/.style = {
        treenode,
        minimum width=9mm,
    },
    smallbasisnode/.style = {
        basisnode, smalltext
    },
    leafnode/.style = {
        align=center,
        text centered,
        font=\footnotesize,
        circle,
        draw=black,
        fill=gray,
        minimum width=1em,
        minimum height=1em,
        inner sep=0mm,
        outer sep=0mm
    },
    inactiveleafnode/.style = {
        leafnode,
        draw=gray,
        text=gray,
        fill=lightgray
    }
}

\pgfmathsetseed{\number\pdfrandomseed}

\usepackage{cleveref}
\Crefname{figure}{Fig.}{Figs.}

\usepackage{hyperref}

\hypersetup{pdftitle={Concepts for Composing Finite Element Function Space Bases},
            pdfauthor={Christian Engwer, Carsten Gräser, Steffen
              Müthing, Simon Praetorius, and Oliver Sander} }

\usepackage{attachfile2}
\usepackage{fancyvrb}

\definecolor{interfacecolor}{rgb}{0.95,0.95,1}
\lstdefinestyle{Example}{}
\lstdefinestyle{Interface}{backgroundcolor=\color{interfacecolor},frame=single}

\newcommand{\cpp}[1]{\lstinline[basicstyle=\ttfamily]!#1!}

\newcommand{\len}[1]{\ensuremath{|#1|}}
\renewcommand{\len}[1]{\ensuremath{\operatorname{len}(#1)}}

\usepackage{changepage}

\newsavebox{\codebox}
\newenvironment{cppexample}{%
    \begin{adjustwidth}{-2em}{}\vspace{-2em}\begin{equation}}
    {\end{equation}\vspace{-1em}\end{adjustwidth}}

\newenvironment{cppexample*}{%
    \begin{adjustwidth}{-2em}{}\vspace{-2em}\begin{equation*}}
    {\end{equation*}\vspace{-1em}\end{adjustwidth}}

\newtheorem{definition}{Definition}
\newtheorem{example}{Example}

\newcommand{\R}{\mathbb{R}}
\newcommand{\N}{\mathbb{N}}
\newcommand{\abs}[1]{{\lvert#1\rvert}}

\newcommand{\op}[1]{\operatorname{#1}}
\newcommand{\st}{\; : \;}

\newcommand{\dune}{
  {\textsc{Dune}}\xspace}

\newcommand{\dunemodule}[1]{
  {\texttt{#1}}}

\definecolor{lightblue}{HTML}{55AAFF}

\graphicspath{{gfx/}}

\title{Concepts for Composing Finite Element Function Space Bases}
\author*[1]{\fnm{Christian} \sur{Engwer}
  \orcid{0000-0002-6041-8228}}\email{christian.engwer@uni-muenster.de}
\affil*[1]{\orgdiv{Institute for Analysis and Numerics},
  \orgname{University of Münster},
  \orgaddress{\street{Einsteinstr. 62}, \postcode{48149} \city{Münster}, \country{Germany}}}

\author[2]{\fnm{Carsten} \sur{Gräser}
  \orcid{0000-0003-4855-8655}}
\affil[2]{\orgdiv{Department Mathematik}, \orgname{Friedrich-Alexander-Universität Erlangen-Nürnberg}, \orgaddress{\street{Cauerstraße 11}, \postcode{91058} \city{Erlangen}, \country{Germany}}}

\author[3]{\fnm{Steffen} \sur{Müthing}}
\affil[3]{\orgdiv{Interdisciplinary Center for Scientific Computing}, \orgname{Heidelberg University}, \orgaddress{\street{Im Neuenheimer Feld 205}, \postcode{69120} \city{Heidelberg}, \country{Germany}}}

\author[4]{\fnm{Simon} \sur{Praetorius} \orcid{0000-0002-1372-4708}}
\affil[4]{\orgdiv{Institut für Wissenschaftliches Rechnen}, \orgname{Technische Universität Dresden}, \orgaddress{\postcode{01062} \city{Dresden}, \country{Germany}}}

\author[5]{\fnm{Oliver} \sur{Sander} \orcid{0000-0003-1093-6374}}
\affil[5]{\orgdiv{Institut für Numerische Mathematik}, \orgname{Technische Universität Dresden}, \orgaddress{\postcode{01062} \city{Dresden}, \country{Germany}}}

\abstract{Finite Element discretizations of coupled multi-physics
  partial differential equation models require the handling of composed
  function spaces. In this paper we discuss software concepts and
  abstractions to handle the composition of
  function spaces, based on a representation of
  product spaces as trees of simpler bases.
  From this description, many different numberings of degrees of freedom by multi-indices can be
  derived in a natural way, allowing to adapt the function spaces to
  very different data layouts, so that it opens the possibility to directly
  use the finite element code with very different linear algebra codes,
  different data structures, and different algebraic solvers.

  A recurring example throughout the paper is the stationary Stokes
  equation with Taylor--Hood elements as these are naturally formulated
  as product spaces and highlight why different storage patterns are
  desirable.

  In the second half of the paper we discuss a particular realization
  of most of these concepts in the \dunemodule{dune-functions} module,
  as part of the \dune{} ecosystem.
}


\begin{document}

\maketitle

\section{Introduction}
\label{sec:intro}

Partial differential equations (PDEs) play a major role in modelling and
simulation of problems in engineering and natural sciences. Together
with the ever growing complexity of these models, simulating these
systems often goes beyond the scope of a single researcher or a single
team.

These are reason why in the last decades different open-source PDE
frameworks were developed, usually building on Finite Element methods
(FEM). Nowadays they provide a rich body of functionality to experiment with new numerical methods or to adapt the code to simulate new and more complex problems.
Successful examples are software packages, such as deal.II~\cite{deal.ii}, FEniCS~\cite{fenics}, FreeFEM++~\cite{freefem}, MFEM~\cite{mfem}, NGSolve~\cite{ngsolve}, upon many others.

A central motivation for such frameworks is to create more complex
simulations from simpler building blocks, e.g., for coupled systems or
multi-physics problems, e.g., Stokes--Darcy coupling, fluid--structure
interaction, reaction--diffusion systems, and many more. These lead naturally to
composed function spaces, where the actual PDEs are formulated on.

While one usually talks about function spaces and discrete subspaces,
the important abstraction for implementing a FEM discretization is actually the {\em
  basis} of the function space.  This is because even though finite
element spaces play a central role in theoretical considerations of
the finite element method, actual computations use coefficient
vectors, which are defined with respect to a particular basis.  Also,
for various finite element spaces, more than one basis is used in
practice.  For example, the space of second-order Lagrangian finite
elements is used both with the nodal (Lagrange)
basis~\cite{braess:2013}, and with the hierarchical
basis~\cite{bank:1996}.  Discontinuous Galerkin spaces can be
described in terms of Lagrange bases, monomial bases, Legendre bases,
and more~\cite{hesthaven_warburton:2008}.  It is therefore important
to be able to distinguish these different representations of the same
space in the application code.

As mentioned, often a method is then not only formulated using a single function
space, but it consists of different physical fields or otherwise uses
more basic function spaces to compose the actual approximation
space. An example, which we will also pick up several times through
out the paper, are the composite space for Navier--Stokes
discretizations, using discrete subspaces of $(H^1)^d\times L^2$.
In the analysis of PDEs and PDE discretizations this compositional
structure shows up in many places. Similarly for PDE software it is
useful to compose more complex spaces and their respective bases.
In particular, vector-valued and mixed finite element
spaces can be written as products of simpler spaces.

\emph{The main focus} of this paper is therefor to discuss concepts and  software abstractions for the
construction of such composed function spaces, the corresponding
composed FEM bases, and how to algebraically represent discrete functions in
these spaces.
Subsets of the concepts presented in this paper can be found in
different state-of-the-art Finite Element software packages.
For example FreeFEM++ supports scalar, vector and
composite
spaces, e.g.
\begin{lstlisting}[style=Example]
  fespace Uh(mesh,[P2,P2]); // Velocity space
  fespace Ph(mesh,P1);      // Pressure space
  fespace Xh=Uh*Ph;         // Stokes space
\end{lstlisting}
The \dune{} module \dunemodule{dune-pdelab} support
a hierarchic tree-like construction of function spaces (and their basis)
for a long time already, c.f. \cite{muething:2015},
and the deal.II library allows to construct composite spaces as a
Cartesian product, implemented by the tuple type C++ class
\cpp{FESystem<dim>}, e.g.
\begin{lstlisting}[style=Example]
  FESystem<dim> fe(FE_Q<dim>(degree + 1) ^ dim, FE_Q<dim>(degree));
\end{lstlisting}
This uses C++11 style ``variadic'' constructors, implementing
overloads for different numbers of components.

Such composed finite element function space bases frequently exhibit a fair amount
of structure. Even more, they often have a natural structure as a
tree. This tree construction of finite element spaces has
first been systematically worked out in~\cite{muething:2015}.

\textbf{Outline:} The paper is organized as follows. In
\Cref{sec:finite_element_trees} we formally define the
construction of trees of discrete function spaces and their respective
bases. We discuss how these concepts of trees carry over to different
numberings of basis functions, which induce a variety of indexing
schemes for storing degrees of freedom in flat or nested data
structures. \Cref{sec:dunefunctions} introduces the
\dunemodule{dune-functions} module and discusses how these abstract concepts are implemented in
\dune{}.
The \dunemodule{dune-functions} module allows to
systematically construct such basis tree and offers a rich body of
functionality to actually work with these in a simulation
code. Additional infrastructure that extends the basis concepts to
improve usage and code reuse concludes the discussion about the implementation.
The complete code is already available as part of the current
\dunemodule{dune-functions} release or will be part of the upcoming
2.11 release.

\section{Function space bases}
\label{sec:finite_element_trees}

We begin by introducing the core concepts for representing function space bases.
The composition of elementary function space bases naturally induces an abstract tree structure.
This structure provides significant flexibility in choosing among various isomorphic representations of the basis.

\subsection{Trees of function spaces}

Throughout this paper, we assume a single fixed domain~$\Omega$, on which all considered function spaces are defined.
While our focus is on function spaces consisting of piecewise polynomial functions with respect to a computational grid, this restriction is not strictly required at this stage.

For a set $R$, we denote by $R^\Omega \colonequals \{f:\Omega \to R\}$
the set of all functions mapping from $\Omega$ to $R$. For domains $\Omega\subset \R^d$,
we write $P_k(\Omega) \subset \R^\Omega$ for the space of all
scalar-valued continuous piecewise polynomials of degree at most $k$ on $\Omega$
with respect to some given triangulation.
We will omit the domain if it can be inferred from the context.

Considering the different finite element spaces that appear in the literature, there are some that we will
call \emph{irreducible}. By this term we mean all bases of scalar-valued functions,
but also others like the Raviart--Thomas basis that cannot easily be written as a combination
of simpler bases.
Many other finite element spaces arise naturally as a combination of simpler ones.
There are primarily two ways how two vector spaces $V$ and $W$ can be combined
to form a new one: sums and products.%
\footnote{While these are also called
internal and external sums, respectively, we stick to the terminology
\emph{sum} and \emph{product} in the following.
}

For sums, both spaces need to
have the same range space $R$, and thus both need to be subspaces of $R^\Omega$.
Then the vector space sum
\begin{equation*}
  V + W
  \colonequals
  \{ v + w \; : \; v \in V, \; w \in W \}
\end{equation*}
in $R^\Omega$ will have that same range space.
For example, a $P_2$-space
can be viewed as a $P_1$-space plus a hierarchical extension spanned by bubble functions~\cite{bank:1996}.
XFEM spaces~\cite{moes_dolbow_belytschko:1999} are constructed by adding particular weighted Heaviside
functions to a basic space to capture free discontinuities.
Such sums
of finite element bases are not the scope of this paper.

The second way to construct finite element spaces from simpler ones uses Cartesian products.
Let $V \subset (\R^{r_1})^\Omega$ and $W \subset (\R^{r_2})^\Omega$ be two function spaces.
Then we define the product of $V$ and $W$ as
\begin{align*}
  V \times W
    \colonequals \big\{ (v,w) \st v \in V, \; w \in W \big\}.
\end{align*}
Functions from this space take values in $\R^{r_1} \times \R^{r_2} = \R^{r_1 + r_2}$.
It should be noted that the Cartesian product of
vector spaces must not be confused with the tensor product of these spaces.
Rather, the $k$-th power
of a single space can be viewed as the tensor product of that space with $\R^k$, i.e,
\begin{align*}
    (V)^k
    = \underbrace{V \times \dots \times V}_{k-\text{times}}
    = \R^k \otimes V.
\end{align*}

The product operation allows to build vector-valued and mixed finite element spaces of arbitrary complexity.
For example, the space of
first-order Lagrangian finite elements with values in $\R^3$ can be seen as the product $P_1 \times P_1 \times P_1$.
The lowest-order Taylor--Hood element is the product $P_2 \times P_2 \times P_2 \times P_1$
of $P_2 \times P_2 \times P_2$ for the velocities with $P_1$ for the pressure.
More factor bases can be included easily, if necessary.  We call such products of
spaces \emph{composite spaces}.

\begin{figure}
    \begin{center}
        \begin{tikzpicture}[
                level/.style={
                    sibling distance = (3-#1)*2cm + 1cm,
                    level distance = 1.5cm
                },
                scale=0.8
            ]
            \node [treenode] {$(P_2\times P_2 \times P_2) \times P_1$}
                child{ node [treenode] {$P_2 \times P_2 \times P_2$}
                    child{ node [treenode] {$P_2$} }
                    child{ node [treenode] {$P_2$} }
                    child{ node [treenode] {$P_2$} }
                }
                child{ node [treenode] {$P_1$} };
        \end{tikzpicture}
    \end{center}
    \caption{Function space tree of the Taylor--Hood space $(P_2 \times P_2 \times P_2)\times P_1$}
    \label{fig:taylor_hood_space_tree}
\end{figure}
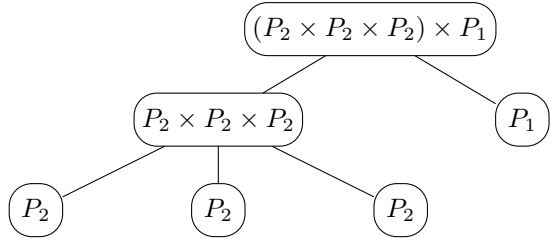

In the Taylor--Hood space, the triple
$P_2 \times P_2 \times P_2$ forms a semantic unit---it contains the components of a velocity field.
The associativity of the product allows to write the Taylor--Hood space
as $(P_2 \times P_2 \times P_2) \times P_1$, which makes the semantic relationship clearer.
Grouped expressions of this type can be conveniently visualized as hierarchical trees, where structural relationships between components become explicit.
This motivates the interpretation of composite finite element spaces as such trees, with leaf nodes representing scalar or otherwise irreducible spaces, and inner nodes denoting products of their children.
Subtrees then represent composite finite element spaces.
\Cref{fig:taylor_hood_space_tree} shows the Taylor--Hood finite element space in such a tree representation.
Note that in this document all trees are \emph{rooted} and \emph{ordered},
i.e., they have a dedicated root node, and the children of each node have a fixed given ordering.
Based on this child ordering we associate to each child the corresponding zero-based index.
Furthermore, we deliberately distinguish nested products of the form
$(V_1 \times \dots \times V_n) \times (W_1 \times \dots \times W_m)$
from flat products
$V_1 \times \dots \times V_n \times W_1 \times \dots \times W_m$.

While the inner tree nodes may initially appear like useless artifacts of the tree representation, they are often extremely useful
because we can treat the sub-trees rooted in those nodes as individual trees in their own right.
This often allows to
reuse existing algorithms that expect to operate on those sub-trees in more complex settings.

\subsection{Trees of function space bases}
\label{sec:basistree}

The multiplication of finite-dimensional spaces naturally induces a corresponding operation on bases
of such spaces.  We introduce a generalized tensor product notation:
Consider linear ranges $R_0,\dots,R_{m-1}$ of function spaces $R_0^\Omega,\dots,R_{m-1}^\Omega$,
and the $i$-th canonical basis vector $\mathbf{e}_i$ in $\R^m$.
Then
\begin{align*}
  \mathbf{e}_i \otimes f
  \colonequals (0,\dots,0,\underbrace{\!f\!}_{\text{\clap{$i$-th entry}}},0,\dots,0)
  \in \prod_{j=0}^{m-1} \Bigl(R_j^\Omega\Bigr) = \Bigl(\prod_{j=0}^{m-1} R_j\Bigr)^\Omega,
\end{align*}
where $0$ in the $j$-th position denotes the zero function in $R_j^\Omega$.
Let $\Lambda_i$ be a (finite) function space basis of the space $V_i = \operatorname{span} \Lambda_i$
for $i=0,\dots,m-1$. Then, a natural basis $\Lambda$ of the product space
\begin{align*}
  V_0 \times \dots \times V_{m-1}
  = \prod_{i=0}^{m-1} V_i
  = \prod_{i=0}^{m-1} \operatorname{span}\Lambda_i
\end{align*}
is given by
\begin{align}
  \label{eq:basis_product}
  \Lambda =
    \Lambda_0 \sqcup \dots \sqcup \Lambda_{m-1}
    = \bigsqcup_{i=0}^{m-1} \Lambda_i
    \colonequals \bigcup_{i=0}^{m-1} \mathbf{e}_i \otimes \Lambda_i.
\end{align}
The product $\mathbf{e}_i \otimes \Lambda_i$ is to be understood element-wise,
and the ``disjoint union'' symbol $\sqcup$ is used here
as a simple short-hand notation for \eqref{eq:basis_product}
and not to be understood as an actual binary operation.
Using this new notation we have
\begin{align*}
  \operatorname{span} \Lambda
    = \operatorname{span} \bigl( \Lambda_0 \sqcup \dots \sqcup \Lambda_{m-1} \bigr)
    = (\operatorname{span} \Lambda_0) \times \dots \times (\operatorname{span} \Lambda_{m-1}).
\end{align*}

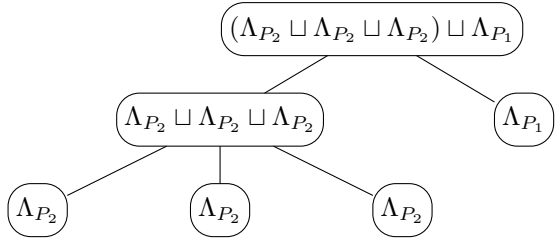
\begin{figure}
    \begin{center}
        \begin{tikzpicture}[
                level/.style={
                    sibling distance = (3-#1)*2cm + 1cm,
                    level distance = 1.5cm
                },
                scale=0.8
            ]
            \node [treenode] {$(\Lambda_{P_2} \sqcup \Lambda_{P_2} \sqcup \Lambda_{P_2}) \sqcup  \Lambda_{P_1}$}
                child{ node [treenode] {$\Lambda_{P_2} \sqcup \Lambda_{P_2} \sqcup \Lambda_{P_2}$}
                    child{ node [treenode] {$\Lambda_{P_2}$} }
                    child{ node [treenode] {$\Lambda_{P_2}$} }
                    child{ node [treenode] {$\Lambda_{P_2}$} }
                }
                child{ node [treenode] {$\Lambda_{P_1}$} };
        \end{tikzpicture}
    \end{center}
    \caption{Function space basis tree of the Taylor--Hood space $(P_2 \times P_2 \times P_2)\times P_1$}
    \label{fig:taylor_hood_basis_tree}
\end{figure}

Similarly to the case of function spaces, bases can be interpreted as trees.
If we associate a basis $\Lambda_V$ to each space $V$ in the function space tree,
then the induced natural function space basis tree is obtained by simply replacing
$V$ by $\Lambda_V$ in each node. For the Taylor--Hood basis this leads to the
tree depicted in \Cref{fig:taylor_hood_basis_tree}.

\subsection{Indexing basis functions by multi-indices}
\label{sec:index_trees}

To work with the basis of a finite element space, the basis functions need to be indexed.  Indexing the basis functions
is what allows to address the corresponding vector and matrix coefficients in suitable vector and matrix data structures.
In simple cases, indexing means simply enumerating the basis functions with natural numbers, but for many applications
hierarchically structured matrix and vector data structures are more natural or efficient.  This leads to the idea
of hierarchically structured multi-indices.
\begin{definition}[Multi-indices]
 A tuple $I \in \N_0^k$ for some $k \in \N_0$ is called a multi-index of length $k$,
 and we write $\len I \colonequals k$.
 The set of all multi-indices is denoted by
 $\mathcal{N} = \bigcup_{k \in \N_0} \N_0^k$.
\end{definition}
To establish some structure in a set of multi-indices it is convenient to consider prefixes.
\begin{definition}[Multi-index prefixes]\mbox{}  
    \begin{enumerate}
        \item
            If $I \in \mathcal{N}$ takes the form $I = (I^0,I^1)$ for $I^0,I^1 \in \mathcal{N}$,
            then we call $I^0$ a prefix of $I$.
            If additionally $\len{I^1}>0$, then we call $I^0$ a strict prefix of $I$.
        \item
            For $I,I^0 \in \mathcal{N}$ and a set $\mathcal{M} \subset \mathcal{N}$:
            \begin{enumerate}
              \item
                We write $I=(I^0,\dots)$, if $I^0$ is a prefix of $I$,
              \item
                we write $I=(I^0,\bullet,\dots)$, if $I^0$ is a strict prefix of $I$,
              \item
                we write $(I^0,\dots) \in \mathcal{M}$, if $I^0$ is a prefix of some
                $I \in \mathcal{M}$,
              \item
                we write $(I^0,\bullet,\dots) \in \mathcal{M}$, if $I^0$ is a strict prefix of some
                $I \in \mathcal{M}$.
            \end{enumerate}
    \end{enumerate}
\end{definition}

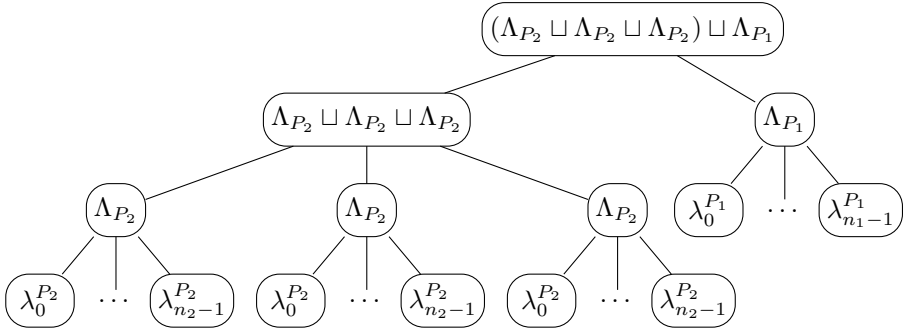
\begin{figure}
    \begin{center}
        \begin{tikzpicture}[
                level/.style={
                    sibling distance = (3-#1)*1.5cm + 1cm,
                    level distance = 1.2cm
                  }
            ]
            \node [treenode] {$(\Lambda_{P_2} \sqcup \Lambda_{P_2} \sqcup \Lambda_{P_2}) \sqcup  \Lambda_{P_1}$}
                child[sibling distance = 7cm]{ node [treenode] {$\Lambda_{P_2} \sqcup \Lambda_{P_2} \sqcup \Lambda_{P_2}$}
                    child [sibling distance = 3.3cm] { node [treenode] {$\Lambda_{P_2}$}
                        child{ node [basisnode] {$\lambda_0^{P_2}$} }
                        child{ node [] {$\dots$} }
                        child{ node [basisnode] {$\lambda_{n_2-1}^{P_2}$} }
                    }
                    child [sibling distance = 3.3cm] { node [treenode] {$\Lambda_{P_2}$}
                        child{ node [basisnode] {$\lambda_0^{P_2}$} }
                        child{ node [] {$\dots$} }
                        child{ node [basisnode] {$\lambda_{n_2-1}^{P_2}$} }
                    }
                    child [sibling distance = 3.3cm] { node [treenode] {$\Lambda_{P_2}$}
                        child{ node [basisnode] {$\lambda_0^{P_2}$} }
                        child{ node [] {$\dots$} }
                        child{ node [basisnode] {$\lambda_{n_2-1}^{P_2}$} }
                    }
                }
                child{ node [treenode] {$\Lambda_{P_1}$}
                    child [sibling distance=1cm] { node [basisnode] {$\lambda_0^{P_1}$} }
                    child [sibling distance=1cm] { node [] {$\dots$} }
                    child [sibling distance=1cm] { node [basisnode] {$\lambda_{n_1-1}^{P_1}$} }
                };
        \end{tikzpicture}
    \end{center}
    \caption{Tree of basis vectors for the Taylor--Hood basis}
    \label{fig:taylor_hood_basis_function_tree}
\end{figure}

It is important to note that the multi-indices from a given set do not necessarily
all have the same length. For an example,
\Cref{fig:taylor_hood_basis_function_tree} illustrates the set of all basis
functions by extending the basis tree of \Cref{fig:taylor_hood_basis_tree}
by leaf nodes for individual basis functions.
A possible indexing of the basis functions of the Taylor--Hood basis $\Lambda_\text{TH}$ then
uses multi-indices of the form $(0,i,j)$ for velocity components, and $(1,k)$
for pressure components.
For the velocity multi-indices $(0,i,j)$, the $i = 0,\dots,2$ determines the component
of the velocity vector field, and the $j = 0,\dots,n_2-1 \colonequals \abs{\Lambda_{P_2}}-1$ determines the number of the scalar $P_2$ basis
function that determines this component.
For the pressure multi-indices $(1,k)$ the $k= 0,\dots,n_1-1 \colonequals \abs{\Lambda_{P_1}}-1$ determines the number of the $P_1$ basis
function for the scalar $P_1$ function that determines the pressure.

It is evident that the complete set of these multi-indices can again be associated to a rooted tree.
In this tree, the multi-indices correspond to the leaf nodes,
their strict prefixes correspond to interior nodes,
and the multi-index digits labeling the edges are the
indices of the children within the ordered tree.  Prefixes can be
interpreted as paths from the root to a given node.

This latter fact can be seen as the defining property of index trees.  Indeed,
a set of multi-indices (together with all its strict prefixes)
forms a tree as long as it is consistent in the sense that the multi-indices
can be viewed as the paths to the leafs in an ordered tree.
That is, the children of each node are enumerated using consecutive zero-based
indices and paths to the leafs (i.e., the multi-indices) are built by concatenating
those indices starting from the root and ending in a leaf.
Since the full structure of this tree is encoded in the multi-indices associated
to the leafs we will---by a slight abuse of notation---call the set of multi-indices
itself a tree from now on.

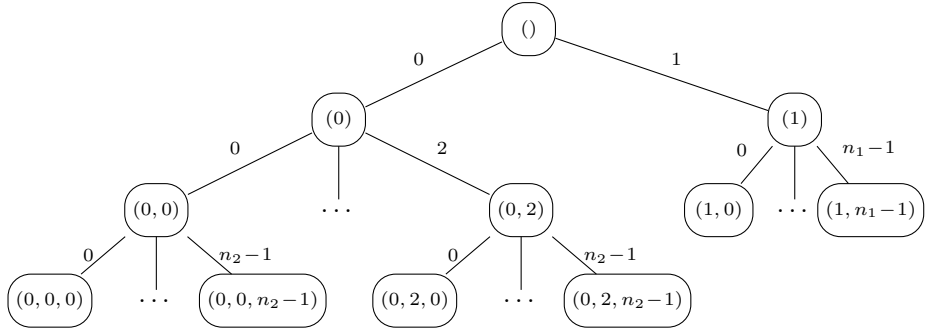
\begin{figure}
  \makebox[\textwidth][c]{
        \begin{tikzpicture}[
                level/.style={
                    sibling distance = (3-#1)*1.5cm + 1cm,
                    level distance = 1.2cm
                },
            ]
            \node [smalltreenode] {$()$}
                child [sibling distance=50mm] { node [smalltreenode] {$( 0 )$}
                    child [sibling distance=24mm] { node [smalltreenode] {$( 0,0 )$}
                        child [sibling distance=14mm] { node [smallbasisnode] {$( 0,0,0 )$} edge from parent node[left,smalltext] {$0$}}
                        child [sibling distance=14mm] { node [] {$\dots$} }
                        child [sibling distance=14mm] { node [smallbasisnode] {$( 0,0,n_2\!-\!1 )$} edge from parent node[right,smalltext] {$n_2\!-\!1$} }
                        edge from parent node[above left,smalltext] {$0$}
                    }
                    child [sibling distance=2.3cm]{ node [] {$\dots$} }
                    child [sibling distance=24mm] { node [smalltreenode] {$( 0,2 )$}
                        child [sibling distance=14mm] { node [smallbasisnode] {$( 0,2,0 )$} edge from parent node[left,smalltext] {$0$}}
                        child [sibling distance=14mm] { node [] {$\dots$} }
                        child [sibling distance=14mm] { node [smallbasisnode] {$( 0,2,n_2\!-\!1 )$} edge from parent node[right,smalltext] {$n_2\!-\!1$} }
                        edge from parent node[above right,smalltext] {$2$}
                    }
                    edge from parent node[above left,smalltext] {$0$}
                }
                child [sibling distance=70mm] { node [smalltreenode] {$( 1 )$}
                    child [sibling distance=1cm] { node [smallbasisnode] {$( 1,0 )$} edge from parent node[above left,smalltext] {$0$} }
                    child [sibling distance=1cm] { node [] {$\dots$} }
                    child [sibling distance=1cm] { node [smallbasisnode] {$( 1,n_1\!-\!1 )$} edge from parent node[above right,smalltext] {$n_1\!-\!1$} }
                    edge from parent node[above right,smalltext] {$1$}
                };
        \end{tikzpicture}
      }
    \caption{Index tree for the Taylor--Hood basis inherited from the basis tree}
    \label{fig:taylor_hood_index_tree}
\end{figure}

\begin{definition}
\label{def:index_tree}
 A set $\mathcal{I} \subset \mathcal{N}$ is called an \emph{index tree}
 if for any $(I,i,\dots) \in \mathcal{I}$ there are also $(I,0,\dots),(I,1,\dots),\dots,(I,i-1,\dots) \in \mathcal{I}$,
 but $I \notin \mathcal{I}$.
\end{definition}
The index tree for the example indexing of the Taylor--Hood basis given above is shown
in \Cref{fig:taylor_hood_index_tree}.

\begin{definition}
Let $(I,\dots) \in \mathcal{I}$, i.e., $I$ is a
prefix of multi-indices in $\mathcal{I}$. Then the size of $\mathcal{I}$ relative
to $I$ is given by
\begin{align}\label{eq:prefix_size}
  \operatorname{deg}^+_{\mathcal{I}}[I] \colonequals  \op{max}\{k \st \exists (I,k,\dots) \in \mathcal{I} \}+1.
\end{align}
\end{definition}
In terms of the ordered tree associated with $\mathcal{I}$ this corresponds
to the out-degree of $I$, i.e., the number of direct children of the node indexed by $I$.

Using the idea of multi-index trees,
an indexing of a function space basis is an injective map from the leaf nodes of a tree of basis functions to the leafs of an
index tree.

\begin{definition}
\label{def:index_map}
  Let $M$ be a finite set and $\iota:M \to \mathcal{N}$ an injective map whose range
  $\iota(M)$ forms an index tree.
  Then $\iota$ is called an \emph{index map} for $M$.
  The index map is called \emph{uniform} if additionally $\iota(M) \subset \mathbb{N}^k_0$ for some $k \in \mathbb{N}$,
  and \emph{flat} if $\iota(M) \subset \mathbb{N}_0$.
\end{definition}

Continuing the Taylor--Hood example, if
all basis functions $\Lambda_\text{TH} = \{\lambda_I \}$ of the whole finite element tree are
indexed by multi-indices of the above given form,
and if $X$ is a coefficient vector that has a compatible hierarchical structure,
then a finite element function $(v_h,p_h)$ with velocity
$v_h$ and pressure $p_h$ defined by the coefficient vector $X$
is given by
\begin{align}
\label{eq:linear_combination}
  (v_h,p_h)
  &= \sum_{i=0}^2\sum_{j=0}^{n_2-1} X_{(0,i,j)}\lambda_{(0,i,j)}
  + \sum_{k=0}^{n_1-1} X_{(1,k)}\lambda_{(1,k)},
\end{align}
with basis functions
\begin{equation*}
  \lambda_{(0,i,j)} = \mathbf{e}_0 \otimes (\mathbf{e}_i \otimes \lambda^{P_2}_j), \qquad i=0,1,2,
    \qquad \text{and} \qquad
    \lambda_{(1,k)} = \mathbf{e}_1 \otimes \lambda^{P_1}_k.
\end{equation*}
Introducing the corresponding index map $\iota : \Lambda_{\text{TH}} \to \mathcal{N}$
with $\iota(\lambda_I)=I$ on the set $\Lambda_{\text{TH}}$ of all basis functions
we can write this in compact form as
\begin{align*}
  (v_h,p_h) &= \sum_{\lambda \in \Lambda_{\text{TH}}} X_{\iota(\lambda)} \lambda
            = \sum_{I \in \iota(\Lambda_{\text{TH}})} X_I \lambda_I.
\end{align*}
Alternatively the individual velocity and pressure fields
$v_h$ and $p_h$ are given by
\begin{align*}
  v_h &= \sum_{i=0}^2 \sum_{j=0}^{n_2-1} X_{(0,i,j)} (\mathbf{e}_i \otimes \lambda^{P_2}_j),
    &
    p_h &= \sum_{k=0}^{n_1-1} X_{(1,k)}\lambda^{P_1}_k.
\end{align*}

\begin{figure}
    \begin{center}
        \begin{tikzpicture}[
                level/.style={
                    sibling distance = (3-#1)*1.5cm + 1.2cm,
                    level distance = 1.2cm
                },
            ]
            \node [treenode] {$()$}
                child{ node [smalltreenode] {$( 0 )$}
                        child [sibling distance=2.5cm] { node [smalltreenode] {$( 0,0 )$}
                            child [sibling distance=1.5cm] { node [smallbasisnode] {$( 0,0,0 )$} edge from parent node[above left,smalltext] {$0$}}
                            child [sibling distance=1.5cm] { node [smallbasisnode] {$( 0,0,1 )$} edge from parent node[left,smalltext] {$1$}}
                            child [sibling distance=1.5cm] { node [smallbasisnode] {$( 0,0,2 )$} edge from parent node[above right,smalltext] {$2$}}
                            edge from parent node[above left,smalltext] {$0$}
                        }
                        child [sibling distance=2.cm]{ node [] {$\dots$} }
                        child [sibling distance=2.8cm]{ node [smalltreenode] {$( 0,n_2\!-\!1 )$}
                            child [sibling distance=2cm] { node [smallbasisnode] {$( 0,n_2\!-\!1,0 )$} edge from parent node[above left,smalltext] {$0$}}
                            child [sibling distance=2cm] { node [smallbasisnode] {$( 0,n_2\!-\!1,1 )$} edge from parent node[left,smalltext] {$1$}}
                            child [sibling distance=2cm] { node
                              [smallbasisnode] {$( 0,n_2\!-\!1,2 )$}
                              edge from parent node[above right,smalltext,yshift=-1mm] {$2$}}
                            edge from parent node[above right,smalltext] {$n_2\!-\!1$}
                        }
                        edge from parent node[above left,smalltext] {$0$}
                }
                child [sibling distance=7.1cm]{ node [smalltreenode] {$( 1 )$}
                    child [sibling distance=1.3cm] { node [smallbasisnode] {$( 1,0 )$} edge from parent node[above left,smalltext] {$0$} }
                    child [sibling distance=1.3cm] { node [] {$\dots$} }
                    child [sibling distance=1.3cm] { node [smallbasisnode] {$( 1,n_1\!-\!1 )$} edge from parent node[above right,smalltext] {$n_1\!-\!1$} }
                    edge from parent node[above right,smalltext] {$1$}
                };
        \end{tikzpicture}
    \end{center}
    \caption{Index tree for Taylor--Hood with blocking of velocity components}
    \label{fig:taylor_hood_index_blocked_tree}
\end{figure}
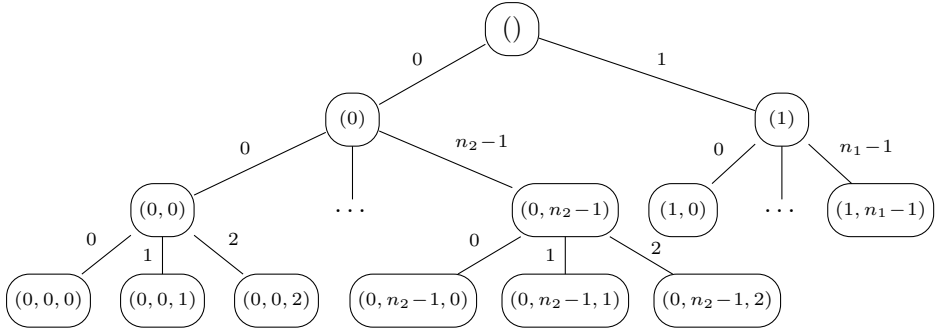

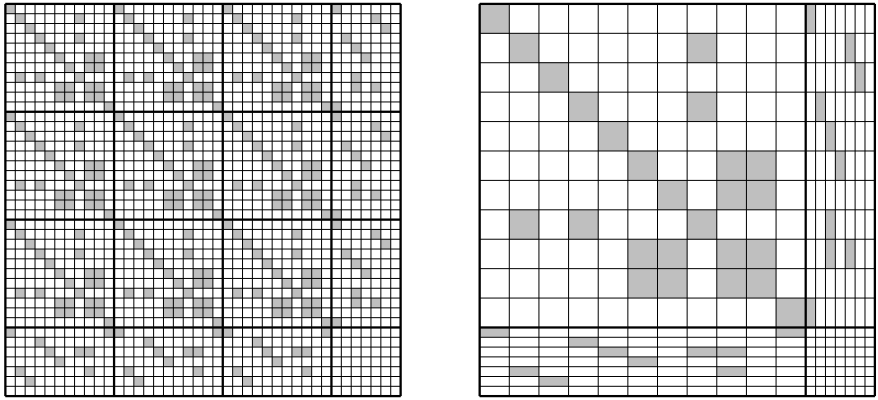
\begin{figure}
 \begin{center}
  \begin{tikzpicture}[scale=.13]

  \pgfmathsetmacro{\offset}{48}


  \foreach \x in {0,...,10}
  {
    \foreach \i in {0,...,2}
      \foreach \j in {0,...,2}
        \fill [lightgray] (11*\i + \x,40- \j*11 - \x) rectangle (11*\i + 1+\x,39 -\j*11 -\x);

    \fill [lightgray] (\offset + 0+3*\x,40-3*\x) rectangle (\offset + 3+3*\x,37-3*\x);
  }

  \foreach \blocknumber in {0,...,9}
  {
    \pgfmathsetmacro{\x}{int(random(0,9))}
    \pgfmathsetmacro{\y}{int(random(\x,9))}

    \foreach \i in {0,...,2}
      \foreach \j in {0,...,2}
      {
        \fill [lightgray] (11*\i + \x,40-11*\j - \y) rectangle (11*\i + 1+\x,39-11*\j-\y);  
        \fill [lightgray] (11*\i + \y,40-11*\j - \x) rectangle (11*\i + 1+\y,39-11*\j-\x);  
      }

    \fill [lightgray] (\offset + 0+3*\x,40-3*\y) rectangle (\offset + 3+3*\x,37-3*\y);  
    \fill [lightgray] (\offset + 0+3*\y,40-3*\x) rectangle (\offset + 3+3*\y,37-3*\x);  
  }

  \foreach \blocknumber in {0,...,9}
  {
    \pgfmathsetmacro{\x}{int(random(0,10))}
    \pgfmathsetmacro{\y}{int(random(0,6))}

    \foreach \i in {0,...,2}
    {
      \fill [lightgray] (11*\i + \x, \y) rectangle (11*\i + 1+\x,1+\y);  
      \fill [lightgray] (40 - \y,40-11*\i - \x) rectangle (39-\y,39-11*\i-\x);  
    }

    \fill [lightgray] (\offset + 0+3*\x,\y)     rectangle (\offset + 3+3*\x,1+\y);  
    \fill [lightgray] (\offset + 40-\y,40-3*\x) rectangle (\offset + 39-\y,37-3*\x);  
  }


  \foreach \x in {0,...,40}
    \draw [line width=0.05mm] (\x,0)--(\x,40);

  \foreach \y in {0,...,40}
    \draw [line width=0.05mm] (0,\y)--(40,\y);

  \foreach \x in {0,11,22,33,40}
    \draw [line width=0.3mm] (\x,0)--(\x,40);

  \foreach \y in {0,7,18,29,40}
    \draw [line width=0.3mm] (0,\y)--(40,\y);


  \foreach \x in {0,...,10}
    \draw [line width=0.05mm] (\offset + 3*\x,0)--(\offset + 3*\x,40);

  \foreach \y in {0,...,10}
    \draw [line width=0.05mm] (\offset + 0,40-3*\y)--(\offset + 40,40-3*\y);

  \foreach \x in {0,...,6}
    \draw [line width=0.05mm] (\offset + 40 - 1*\x,0)--(\offset + 40 - 1*\x,40);

  \foreach \y in {0,...,6}
    \draw [line width=0.05mm] (\offset + 0,\y)--(\offset + 40,\y);

  \foreach \x in {0,33,40}
    \draw [line width=0.3mm] (\offset + \x,0)--(\offset + \x,40);

  \foreach \y in {0,7,40}
    \draw [line width=0.3mm] (\offset + 0,\y)--(\offset + 40,\y);

  \end{tikzpicture}
 \end{center}
 \caption{Two matrix occupation patterns for different indexings of the Taylor--Hood bases.
   Left: Corresponding to the index tree of \Cref{fig:taylor_hood_index_tree}.
   Right: Corresponding to the index tree of \Cref{fig:taylor_hood_index_blocked_tree}.
   }
 \label{fig:matrix_occupation_patterns}
\end{figure}

In the previous example, the index tree was
isomorphic to the basis function tree depicted in \Cref{fig:taylor_hood_basis_function_tree}.
However, one may also be interested in constructing multi-indices
that do not mimic the structure of the basis function tree:
For example, to increase data locality in assembled matrices for the Taylor--Hood basis it may be
preferable to group all velocity degrees of freedom corresponding to a single
$P_2$ basis function together, i.e., to use the index $(0,j,i)$
for the $j$-th $P_2$ basis function for the $i$-th component.
The corresponding alternative index tree is shown in
\Cref{fig:taylor_hood_index_blocked_tree}.
In \Cref{fig:matrix_occupation_patterns} the corresponding layouts of a hierarchical stiffness matrix is visualized.

Alternatively, the case of indexing all basis functions from the Taylor--Hood basis with a single
natural number can be represented by an index tree with $3 n_2 + n_1$ leaf nodes all
directly attached to a single root. Different variations of such a tree differ by how the
degrees of freedom are ordered.

\subsection{Localization to single grid elements}
\label{sec:localization}
A common practice in Finite Element libraries is that that all
important information is accessed element by element; matrices and
vectors are assembled element by element and also a global basis is
often defined element by element, following the concepts of Ciarlet~\cite{ciarlet:1978}.
It is therefore important
to consider the restrictions of bases to single grid elements. We now
require that there is a finite element grid for the domain $\Omega$
and for simplicity we
assume that all bases consist of functions that are defined piecewise with respect to this grid,
but it is actually sufficient to require that the restrictions of all basis functions to elements
of the grid can be constructed cheaply.

Considering the restrictions of the global basis functions $\lambda
\in \Lambda$ of a given tree to a single fixed grid element $e$,
we discard all those that are constant zero functions on $e$.
All others form the \emph{local basis} on $e$
\begin{equation*}
 \Lambda|_e
 \colonequals
 \{ \lambda|_e \; : \; \lambda \in \Lambda,
         \quad \operatorname{int}(\operatorname{supp} \lambda) \cap e \neq \emptyset \}.
\end{equation*}
This local basis forms a tree that is isomorphic to the original function space basis tree,
with each global function space basis $\Lambda$ replaced by its local counterpart $\Lambda|_e$.
For a given index map $\iota$ of $\Lambda$,
this natural isomorphism from global to local tree
naturally induces a localized version of $\iota$ given by
\begin{align*}
  \iota|_e : \Lambda|_e &\to \mathcal{I}, &
  \iota|_e(\lambda_e) &\colonequals \iota(\lambda),
\end{align*}
which is the map that associates shape functions on a given grid element $e$ to
the multi-indices of the corresponding global basis functions.
Note that the map $\iota|_e$ itself is not an index map in the sense of \Cref{def:index_map}
since $\iota|_e(\Lambda|_e)$ is only a subset of the index tree $\iota(\Lambda)$,
and not always an index tree itself.

In order to index the basis functions in $\Lambda|_e$ efficiently we introduce
an additional local index map
\begin{align*}
  \iota^{\text{local}}_{\Lambda|_e}: \Lambda|_e \to
  \mathbb{N}_0.
\end{align*}
The index $\iota^{\text{local}}_{\Lambda|_e}(\lambda|_e)$ is
called the \emph{local index} of $\lambda$ (with respect to $e$).
To distinguish it from the indices generated by $\iota$
we call $\iota(\lambda)$ the \emph{global index} of $\lambda$.
The local index is typically used to address element stiffness
matrices or load vectors.
Note, that in principle, this local indexing could also be defined using an index
tree, but for simplicity and ease of implementation it is commonly a
flat index.

The leaf nodes naturally have a local indexing, as they appear as a
basis of their own. We introduce this numbering of a leaf local basis $\hat{\Lambda}|_e$
as
\begin{align*}
  \iota^{\text{leaf}}_{\hat{\Lambda}|_e}: \hat{\Lambda}|_e \to \mathbb{N}_0,
\end{align*}
called the \emph{leaf-local index} of $\lambda$ (with respect to $e$).

In practice one typically accesses
basis functions by indices directly, e.g., to store the assembled
values into the local load vector.
Hence we additionally define the maps
\begin{align*}
  \iota^{\text{leaf}\to\text{local}}_e \colonequals \iota^{\text{local}}_{\Lambda|_e} \circ (\iota^{\text{leaf}}_{\hat{\Lambda}|_e})^{-1}
\end{align*}
mapping leaf-local indices to local indices and
\begin{align*}
  \iota^{\text{local}\to\text{global}}_e \colonequals \iota|_e \circ (\iota^{\text{local}}_{\Lambda|_e})^{-1}
\end{align*}
mapping local indices to global multi-indices.

\subsection{Strategy-based construction of multi-indices}
\label{sec:index_strategies}
As discussed, for $\Lambda$, the set of basis functions of a finite element basis tree, many different indexing schemes are feasible.
The only requirement is that the indexing map $\iota: \Lambda \to \mathcal{N}$ is injective and that its image forms a valid index tree.
Among the many admissible choices, certain indexing schemes are particularly useful in practice.
These can be constructed systematically using well-defined transformation rules.
The construction proceeds recursively.
To describe it,
we assume in the following that $\Lambda$ is a node in the function space
basis tree, i.e., it is the set of all basis functions
corresponding to a node $V \colonequals \operatorname{span} \Lambda$
in the function space tree.

To end the recursion, we assume that an index map $\iota : \Lambda \to \mathcal{N}$
is given if $V = \operatorname{span} \Lambda$ is a leaf node of the function space tree.
The most obvious choice would be a flat zero-based index
of the basis functions of $\Lambda$. However, other choices are possible.
For example, in case of a discontinuous finite element space, each
basis function $\lambda \in \Lambda$ could also be associated to a two-digit
multi-index $\iota(\lambda)=(i,k)$, where $i$ is the
index of the grid element that forms the support of $\lambda$, and $k$ is the index of $\lambda$
within this element.

For the actual recursion,
if $\Lambda$ is any non-leaf node in the function space basis tree,
then it takes the form
\begin{align*}
  \Lambda = \Lambda_0 \sqcup \dots \sqcup \Lambda_{m-1}
          = \bigcup_{i=0}^{m-1} \mathbf{e}_i \otimes \Lambda_i,
\end{align*}
where $\Lambda_0, \dots,\Lambda_{m-1}$ are the direct children of $\Lambda$,
i.e., the sets of basis functions of the child
spaces $\{\operatorname{span} \Lambda_i\}_{i=0,\dots,m-1}$ of the product space
\begin{align*}
  \operatorname{span} \Lambda
    = \operatorname{span} \bigl( \Lambda_0 \sqcup \dots \sqcup \Lambda_{m-1} \bigr)
    = (\operatorname{span} \Lambda_0) \times \dots \times (\operatorname{span} \Lambda_{m-1}).
\end{align*}
For the recursive construction, we assume that an index map
$\iota_i : \Lambda_i \to \mathcal{N}$ on $\Lambda_i$ is given for any $i=0,\dots,m-1$.
The task is to construct an index map $\iota: \Lambda \to \mathcal{N}$
from the maps $\iota_i$.

Before discussing the strategies, we need to introduce the special
case of a power node, as some strategies are only applicable to these.
\begin{definition}[Power node]
\label{def:power_node}
  An inner node $\Lambda$ will be called \emph{power node} if all of its children $\Lambda_i$
  are identical and equipped with identical index maps $\iota_i$.
  An inner node that is not a power node is called \emph{composite node}.
\end{definition}

\begin{example}[Generic blocking strategies]\label{example:blocked-merging-strategies}
In the following, we describe different strategies that are or have
been used in practice.
Remember that any $\lambda \in \Lambda$ has a unique representation
$\lambda = \mathbf{e}_i \otimes \hat{\lambda}$ for $i \in \{0,\dots,m-1\}$ and some
$\hat{\lambda} \in \Lambda_i$.
\begin{description}[itemsep=.5ex, topsep=.5ex, leftmargin=*]
  \item[\textbf{blocked-lexicographic}:] This strategy prepends the child index
    to the multi-index within the child basis. That is, the index map $\iota:\Lambda \to \mathcal{N}$
    is given by
    \begin{align*}
      \iota(\mathbf{e}_i \otimes \hat{\lambda}) = (i,\iota_i(\hat{\lambda})).
    \end{align*}

  \item[\textbf{blocked-interleaved}:] This strategy is only well-defined for power nodes. It appends the child index
    to the multi-index within the child basis.
    That is, the index map $\iota:\Lambda \to \mathcal{N}$
    is given by
    \begin{align*}
      \iota(\mathbf{e}_i \otimes\hat{\lambda}) = (\iota_i(\hat{\lambda}),i).
    \end{align*}
\end{description}
\end{example}
\begin{example}[Generic flat strategies]\label{example:flat-merging-strategies}
Unlike the previous two strategies, the following two do not introduce new
multi-index digits. Such strategies are called \emph{flat}.
\begin{description}[itemsep=.5ex, topsep=.5ex, leftmargin=*]
  \item[\textbf{flat-lexicographic}:] This strategy merges the roots of all
    index trees $\iota_i(\Lambda_i)$ into a single new one.
    Assume that we split the multi-index
    $\iota_i(\hat{\lambda})$ according to
    \begin{align}\label{eq:multiindex-split}
      \iota_i(\hat{\lambda}) = (i_0,I),
    \end{align}
    where $i_0 \in \mathbb{N}_0$ is the first digit.
    The index map $\iota:\Lambda \to \mathcal{N}$ is then given by
    \begin{align*}
      \iota(\mathbf{e}_i \otimes\hat{\lambda}) = (L_i + i_0, I),
    \end{align*}
    where the offset $L_i$ for the first digit is computed by
    $L_i = \sum_{j=0}^{i-1} \operatorname{deg}_{\iota_j(\Lambda_j)}^+[()]$
    and by this ensures that $\iota$ always forms a consecutive index map for $\Lambda$.

  \item[\textbf{flat-interleaved}:] This strategy again only works for power nodes.
    It also merges
    the roots of all child index trees $\iota_i(\Lambda_i)$
    into a single one, but it interleaves the children.
    Again using the splitting
    $\iota_i(\hat{\lambda}) = (i_0,I)$ introduced in~\eqref{eq:multiindex-split},
    the index map $\iota:\Lambda \to \mathcal{N}$ is given by
    \begin{align*}
      \iota(\mathbf{e}_i \otimes\hat{\lambda}) = (i_0 m + i, I),
    \end{align*}
    where the fixed stride $m$ is given by the number of children of $\Lambda$.
\end{description}
\end{example}

Additional index-merging strategies can often be formulated when more structural information about the underlying index maps~$\iota_i$ is available.
For example, if it is known that basis functions can be associated with grid entities such as vertices, edges, facets, or cells, and that certain digits of the index maps~$\iota_i$ correspond to indices of these entities, then one can define strategies that block all basis functions associated with the same grid entity.

\begin{example}[Entity-wise blocking strategy]
  Assume that every basis function $\hat\lambda \in \Lambda_i$ is
  associated to a grid entity $e_{\hat \lambda}$.
  Furthermore, we assume that the grid entity $e$ has the
  index $I(e)$
  and that the index of $\hat \lambda$ in $\Lambda_i$ is given by
  \begin{align*}
    \iota_i(\hat \lambda) = (I(e_{\hat \lambda}), j_{\hat\lambda})
  \end{align*}
  where
  $j_{\hat \lambda} \in \{0, \dots, n_i(e_{\hat\lambda})-1\}$
  is a flat index
  and $n_i(e) := \operatorname{deg}_{\iota_i(\Lambda_i)}^+[I(e)]$
  is the number of basis functions from $\Lambda_i$
  associated to the grid entity $e$.
  Naturally, $j_{\hat\lambda}$ has to be unique among all basis functions
  in $\Lambda_i$ associated to the same grid entity.
  Then we can construct indices of the basis functions in $\Lambda$
  by the following entity-wise blocking strategy:
\begin{description}[itemsep=.5ex, topsep=.5ex, leftmargin=*]
  \item[\textbf{blocked-by-entity}:]
    Associate to the basis function $\mathbf{e}_i \otimes\hat{\lambda} \in \Lambda$ the index
    \begin{align*}
      \iota(\mathbf{e}_i \otimes\hat{\lambda}) = (I(e_{\hat \lambda}), j_{\hat\lambda}+N_i(e_{\hat \lambda}))
    \end{align*}
    with $N_i(e) = \sum_{j=0}^{i-1} n_i(e)$ denoting the entity local offset.
    Notice that this index map $\iota$ on $\Lambda$ satisfies the
    above assumption, such that the blocked-by-entity strategy
    can be applied recursively.
\end{description}
\end{example}

\section{Realization of composable bases in dune-functions}
\label{sec:dunefunctions}
As part of the \dune ecosystem, the \dunemodule{dune-functions} module
implements major parts of the discussed concepts.
In the following, we refer to the most recent released version 2.10 of the
\dunemodule{dune-functions}\footnote{\url{https://dune-project.org/modules/dune-functions},
gitlab repository \url{https://gitlab.dune-project.org/staging/dune-functions/}}
module, published along with the official \dune{} 2.10 release~\cite{BlattEtAl2025Dune2.10}.
Additionally, we will highlight a few specific changes and extensions
that are currently under development.
Before elaborating on the realization of these composable bases
concepts, we want to briefly introduce the \dune framework.

\subsection{The DUNE ecosystem}
\dune is a C++ software framework and ecosystem focussing on grid-based discretization methods for partial differential equations. It has a modular
structure, so that users can pick what they need. Further it allows
for the development of experimental new features in separate modules
without risking stability of the main components.

The \emph{core modules} of the \dune software system focus on low-level infrastructure for
implementations of simulation algorithms for partial differential equations.  Modules like
\dunemodule{dune-grid}\,\cite{dune08:1,dune08:2} and \dunemodule{dune-istl}\,\cite{BlattBastian2007DuneIstl,BlattBastian2008Generic} provide programmer interfaces (APIs) to finite element grids
and sparse linear algebra, respectively, but little more. Actual finite element functions only
appear in the \dunemodule{dune-localfunctions} module, which deals with discrete function spaces
on single grid elements exclusively.

On top of these core modules, various other modules in the \dune ecosystem implement finite element and finite volume assemblers
and solvers, and the corresponding discrete function spaces. The most prominent ones are
\dunemodule{dune-pdelab} \cite{BlattHeimannMarnach2010Generic,dune-pdelab}
and \dunemodule{dune-fem} \cite{DednerEtAl2010DuneFem,dune-fem},
but smaller ones like \dunemodule{dune-fufem} \cite{dune-fufem}
and \dunemodule{AMDiS} \cite{amdis},
exist as well. The functionality of these modules overlaps to a considerable extent, even though
each such module has a different focus.

The \dunemodule{dune-functions}  module was written to partially overcome this fragmentation,
and to unify parts of the different implementations.

\subsection{The dune-functions module}
The \dunemodule{dune-functions} picks well-defined aspects of finite
element assembly---finite element spaces and functions---and, in the
\dune spirit, provides abstract interfaces that try to be both
flexibly and efficient.

It
provides interfaces for functions and
function space bases. It forms one abstraction level above grids,
shape functions, and linear algebra, and provides infrastructure for
full discretization frameworks like \dunemodule{dune-pdelab} and
\dunemodule{dune-fem}.
Indeed, at the time of writing
the modules  \dunemodule{AMDiS} and \dunemodule{dune-fufem} are based
on \dunemodule{dune-functions} and
\dunemodule{dune-pdelab} is in the process
of migration.

This
document describes the function space bases provided by
dune-functions. These are based on an abstract description of bases
for product spaces as trees of simpler bases. From this description,
many different numberings of degrees of freedom by multi-indices can
be derived in a natural way. We describe the abstract concepts,
document the programmer interface, and give a complete example program
that solves the stationary Stokes equation using Taylor--Hood elements.

Of the two parts of \dunemodule{dune-functions} functionality, the APIs for discrete and
closed-form functions have already been described in a separate paper~\cite{engwer_graeser_muething_sander:2015}.

\begin{figure}
 \begin{center}
  \includegraphics[width=0.9\textwidth]{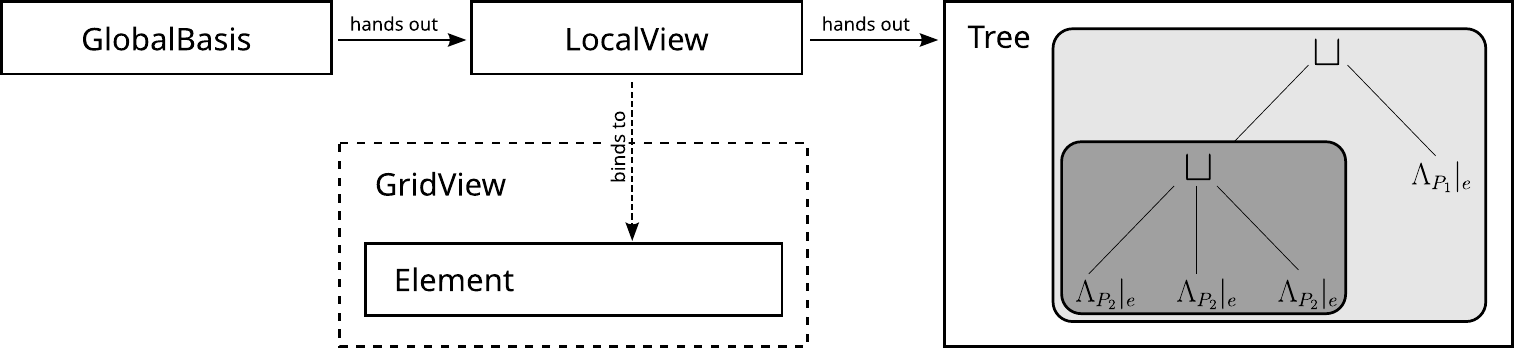}
 \end{center}
 \caption{Overview of the classes making up the interface to finite element space bases}
 \label{fig:febasis_interface_schematic}
\end{figure}

For these reasons, the main \dunemodule{dune-functions} interface represents a basis of a
discrete function space, and not the space itself.
The design of these function spaces bases
follows the ideas of the previous section. The main interface concept are global basis objects
that represent trees of function space bases. These trees can be localized to individual elements
of the grid. Such a localization provides access to the (tree of) shape functions there,
together with the two shape-function index maps
$\iota^{\text{leaf}\to\text{local}}_e$ and
$\iota^{\text{local}\to\text{global}}_e$.
The structure of the interface is visualized in \Cref{fig:febasis_interface_schematic}.

\subsection{Construction and usage of discrete function space bases}
\label{sec:functions-basis-construction}
In \dunemodule{dune-functions}
a tree of global bases is implemented by a class which,
in the following, we will call \cpp{GlobalBasis}.
It provides access to the basis tree, providing functionality to
localize the basis and extract information about the basis and its
index tree.
Internally, different implementations provide the functionality of
different leaf basis nodes, composite nodes or power nodes.
In \dunemodule{dune-functions}, construction of the actual basis is
usually done via a ``factory'' function:
\begin{lstlisting}[style=Interface]
template<class GV, class F>
auto makeBasis(const GV& gridView, const F& factory);
\end{lstlisting}
where the \cpp{factory} argument provides information about the
individual nodes in the tree and its nesting structure.

We start with an example of how to define the tree structure of a Taylor--Hood basis, with a blocking of indices corresponding to the index-tree structure illustrated in \Cref{fig:taylor_hood_index_blocked_tree}:
\begin{lrbox}{\codebox}
\begin{lstlisting}[style=Example]
using namespace Dune;
using namespace Functions::BasisFactory;

auto taylorHoodBasis = makeBasis(gridView,
  composite(
    power<dim>(
      lagrange<p+1>(),
      blockedInterleaved()),
    lagrange<p>(),
    blockedLexicographic()
  ));
\end{lstlisting}
\end{lrbox}
\begin{cppexample}\usebox{\codebox}\label{ex:taylor-hood-basis}\end{cppexample}
The \cpp{lagrange<p>()} function creates an object describing a Lagrange function space
with a statically defined polynomial order\footnote{Similarly, the function \cpp{lagrange(., p)} describes a Lagrange function spaces with dynamic order.}, \cpp{power<dim>}\footnote{Similarly,  the function \cpp{power(., dim)} represents a power node with dynamic degree.}, and \cpp{composite} respectively are factories to create the descriptors for power and composite nodes.

In addition to the \cpp{lagrange<p>()} factory, the \dunemodule{dune-functions} module provides support for a broad range of irreducible finite element basis implementations. Most of these are constructed on top of locally defined bases supplied by the \dunemodule{dune-localfunctions} module, although non-standard bases---such as B-spline bases---are also supported. The implementation encompasses a variety of function spaces, including:
$H^1$-conforming bases (e.g., Lagrange and hierarchical Lagrange),
$H^{\operatorname{div}}$- and $H^{\operatorname{curl}}$-conforming bases (e.g., Raviart--Thomas, Brezzi--Douglas--Marini, N\'ed\'elec),
non-conforming and discontinuous bases (e.g., Rannacher--Turek, Morley, Lagrange--DG), and
$H^2$-conforming bases (e.g., cubic Hermite).

The nodes in the example above also contain additional meta-data, which contains information about the strategy for multi-index construction.
This will be called \emph{merging strategy} in the following.
In \dunemodule{dune-functions} we directly support blocked-lexicographic, blocked-interleaved, flat-lexicographic, and flat-interleaved strategies, see \Cref{example:blocked-merging-strategies,example:flat-merging-strategies} for their abstract definition.
We use camel case notation, e.g., \cpp{FlatLexicographic}, to denote the corresponding tags used in the factories.
A particular feature of these four types is that it is possible to generate the final multi-index on the fly by passing index information upwards through the tree.
The merging strategies specified in the example are
\cpp{BlockedLexicographic} for the composite node, and
\cpp{BlockedInterleaved} for the power node. Other merging strategies
can be implemented via a well-defined developers interface.
Note that the interfaces in \dunemodule{dune-functions} also support the
implementation of a \emph{blocked-by-entity} numbering, but the currently
generic construction does not support the automatic construction of
these multi-indices.

\subsubsection{The GlobalBasis}\label{subsec:global-basis}
{The \cpp{GlobalBasis}}, in the example the \cpp{taylorHoodBasis}, has
two main features: giving access to basis functions and their indices.
Most of this access happens through the localization of the basis to single grid elements (\Cref{sec:localization}).
In the programmer interface, this localization is called \cpp{LocalView}.  Objects of type
\cpp{LocalView} are obtained from \cpp{GlobalBasis} objects through the method
\begin{lstlisting}[style=Interface]
using LocalView = @@<implementation defined>@@;
LocalView localView() const;
\end{lstlisting}
The precise return type of the \cpp{localView} method is implementation-dependent.
Objects created by the method have a mostly undefined state until they are attached to individual grid elements in a process called \emph{binding}.
Only a few properties that do not depend on the grid elements can be used in the unbound state.

Additionally, it provides the size information necessary to manage the containers that store the associated coefficients. \Cref{sec:global-basis-and-coefficients} describes how to combine a global basis with coefficient vectors.

\subsubsection{LocalView of a basis}\label{subsec:localview}
The \cpp{LocalView} represents the localization of a function space basis to a single grid element
and encapsulates the local information as described in \Cref{sec:localization}.
A freshly constructed \cpp{LocalView} object is not completely initialized yet.
To truly have the object represent the basis localization on a particular element,
it must be \emph{bound} to that element.  This is achieved by calling
\begin{lstlisting}[style=Interface]
using GridView = typename GlobalBasis::GridView;
using Element = typename GridView::template Codim<0>::Entity;
void LocalView::bind(const Element& e);
\end{lstlisting}
Once this method has been called, the \cpp{LocalView} object is fully set up
and can be used.
The call may incorporate expensive calculations needed to
pre-compute the local basis functions and their global indices.
The local view can be
bound to another element at any time by calling the \cpp{bind} method again.

This allows to iterate through all elements and access the local basis information
by binding the \cpp{LocalView}:
\begin{lstlisting}[style=Example]
// A view on the basis basis on a single element
auto localView = taylorHoodBasis.localView();

// Loop over all grid elements
for (const auto& e : elements(taylorHoodBasis.gridView()))
{
  // Bind the local FE basis view to the current element
  localView.bind(e);
}
\end{lstlisting}

Access to the actual local basis functions is provided by the method
\begin{lstlisting}[style=Interface]
using Tree = @@<implementation defined>@@;
const Tree& LocalView::tree() const;
\end{lstlisting}
This encapsulates the set~$\Lambda|_e$ of basis functions restricted to the element~$e$, organized as a tree structure reflecting the composition of the function space basis.
While the tree can be accessed even when the local view is unbound, it must be bound to a specific element to use most of its functionality.

Traversing \cpp{localView.tree()} provides access to the leaf nodes, which contain the actual local shape functions along with important meta-data, such as the corresponding global multi-indices.

To access the associated local finite element, the leaf nodes provide the method
\begin{lstlisting}[style=Interface]
using FiniteElement = @@<implementation defined>@@;
const FiniteElement& LocalView::finiteElement() const;
\end{lstlisting}
The returned reference points to a \cpp{LocalFiniteElement} that conforms to the interface defined in the \dunemodule{dune-localfunctions} module.
As such, it provides access to shape function values and their derivatives, to the evaluation of degrees of freedom in the sense of~\cite{ciarlet:1978}, and to the assignment of local degrees of freedom to sub-entities of the element.

The numbering of shape functions used by \dunemodule{dune-localfunctions} coincides with the leaf-local indices introduced in \Cref{sec:localization}.

Further the bound \cpp{LocalView} object provides information
about the size of the local basis at the current element and the local
numbering of the shape functions in all leaf nodes.
The total number of basis functions associated to the
local view at the current element is returned by
\begin{lstlisting}[style=Interface]
std::size_t LocalView::size() const;
\end{lstlisting}
In the language of \Cref{sec:finite_element_trees}, this method computes
the number $\abs{\Lambda|_e}$.

\subsubsection{Global numbering}\label{subsec:localview-index}
For any of the local basis functions in the local tree
accessible by \cpp{tree()} the global multi-index
is provided by the method
\begin{lstlisting}[style=Interface]
using MultiIndex = @@<implementation defined>@@;
MultiIndex LocalView::index(std::size_t i) const;
\end{lstlisting}
The argument for this method is the local
index of the basis function within the tree as
returned by \cpp{node.localIndex(k)};
here \cpp{node} is a leaf node of the
tree provided by \cpp{tree()}, and \cpp{k}
is the number of the shape function within the corresponding
local finite element (see below).
Hence the \cpp{index} method
implements the map $\iota^{\text{local}\to\text{global}}_e$
introduced in \Cref{sec:localization},
which maps local indices to global multi-indices.
Accessing the same global index multiple times
is expected to be cheap, because implementations are supposed to pre-compute
and cache indices during \cpp{bind(Element)}.
The result of calling \cpp{index(std::size_t)} in
unbound state is undefined.

Extending the previous example a little, the following loop prints the
global indices for each degree of freedom of each element.
\begin{lstlisting}[style=example]
auto localView = basis.localView();

for (const auto& element : elements(basis.gridView()))
{
  localView.bind(element);
  for (std::size_t i = 0; i < localView.size(); ++i)
    std::cout << localView.index(i) << std::endl;
}
\end{lstlisting}
When this code is run for a Taylor--Hood basis on a two-dimensional triangle grid,
it will print 15 multi-indices per element, because a Taylor--Hood element has 12
velocity degrees of freedom and 3 pressure degrees of freedom per triangle.

\subsection{Implementation and efficiency considerations}\label{subsec:performance}

As mentioned, a central design goal was to balance flexibility and efficiency. For low-order methods, evaluating local shape functions performs a few floating-point operations (FLOPs), so using dynamic polymorphism to implement the \cpp{LocalView::Tree}, e.g., by using abstract
interfaces or type-erasure, would significantly impact performance.

For this reason \dunemodule{dune-functions}
deliberately keeps the types of all nodes explicit, allowing the
compiler to apply all available optimizations. As a consequence, the
tree is not only a tree of objects, but it is also a tree of C++
types, implemented with the utilities offered by the
\dunemodule{dune-typetree} module.
In particular, the composite node consists
of children of different type, technically comparable to
\cpp{std::tuple<children...>}.

A particular consequence of using explicit types in the tree is
that identifying nodes or basis functions (see
\Cref{sec:index_trees}) and choosing a suitable multi-index
representation can be difficult.
One idea would be to use C++ containers, such as \cpp{std::vector<unsigned int>} or \cpp{std::array<unsigned int, N>}. However, this is not feasible because the stored index would be a dynamic information. Accessing a child of a composite node via its index requires compile-time information. For example, this information could be represented by a \cpp{std::integral_constant} instead of an \cpp{unsigned int}.

To overcome this limitation, \dunemodule{dune-functions} uses a special multi-index data type, \cpp{TypeTree::HybridTreePath}\footnote{A \cpp{HybridTreePath} is
similar to a \cpp{std::tuple} of \cpp{std::size_t} and
\cpp{std::integral_constant<std::size_t,.>}.}\footnote{For 2.11 a generalization
\cpp{HybridMultiIndex} in \dunemodule{dune-common} is in development.} from the \dunemodule{dune-typetree} module.
This data type allows the mixing static and dynamic index information to identify sub-trees of a composed function space basis.
Power nodes allow access to children using dynamic indices, while composite nodes require static indices.
For example, accessing the nodes for the different velocity-components of the Taylor--Hood basis is achieved by a combination of static and dynamic indices:
\begin{lstlisting}[style=Example]
using namespace Indices;  // Import namespace with index constants _0, _1, _2, etc
for (unsigned int d = 0; d < dim; ++d) {
    auto treePath = Dune::TypeTree::treePath(_0, d);
    const auto& v_component = localView.tree().child(treePath);
    const auto& fe = v_component.finiteElement();
    @@<...>@@
}
\end{lstlisting}
As all velocity-components have the same C++ type, it is
possible to iterate dynamically from 0 to dim, but the distinction
between the velocity sub-tree and the pressure sub-tree works only
with a static index from the \cpp{Indices} namespace.

On the other hand, the identification of degrees of freedom by multi-indices returned by the \cpp{LocalView} does not strictly require the static and dynamic index mixture, since it is mainly used to address entries in nested vectors or matrices, see \Cref{subsec:backends}. Due to the automatic construction of these indices during traversal of the index tree and because the tree might not be balanced, a container with a statically known maximal capacity but dynamic size is appropriate, such as the \cpp{Dune::ReservedVector}.

\subsection{Advanced bases types}\label{subsec:special-basis-nodes}

\subsubsection{Treating sub-trees as separate bases}
\label{subsec:subspace-basis}

\Cref{sec:functions-basis-construction} has shown how trees of bases can be combined to form bigger trees.
It is also possible to extract sub-trees from other trees and treat these sub-trees as basis trees in their own right.
As they still belong to the big tree, they represent a subspace basis of the composite basis.
An important property of these subspace bases is that the numbering of the coefficients is still that of the large basis.

Considering again the Taylor--Hood basis, a good reason to operate on sub-trees could be the construction of a Schur-complement solver or the visualization of a specific physical field.

The programmer interface for such sub-tree bases is called \cpp{SubspaceBasis}.
It mostly coincides with the interface of a global basis, but additionally to the \cpp{GlobalBasis} interface the \cpp{SubspaceBasis} provides information about where the sub-tree is located in the global basis.
More specifically, the method
\begin{lstlisting}[style=Interface]
const @@<implementation defined>@@& SubspaceBasis::rootBasis() const;
\end{lstlisting}
provides access to the root basis, and the method
\begin{lstlisting}[style=Interface]
using PrefixPath = TypeTree::HybridTreePath@@<implementation defined>@@;
const PrefixPath& SubspaceBasis::prefixPath() const;
\end{lstlisting}
returns the path of the sub-tree associated to the \cpp{SubspaceBasis} within the full tree.
For convenience, a global basis behaves like a trivial \cpp{SubspaceBasis}, i.e., it has the method \cpp{rootBasis} returning the basis itself, and \cpp{prefixPath} returning an empty tree-path.
Note that a \cpp{SubspaceBasis} differs from a full global basis because the global multi-indices are the same as the ones of the root basis, and thus they are in general neither consecutive nor zero-based.
Instead, those multi-indices allow to access containers storing coefficients for the full root basis.

\cpp{SubspaceBasis} objects are created using a global factory function from the root basis and the path to the desired sub-tree. The path can either be passed as a single \cpp{HybridTreePath} object, or as a sequence of individual indices.
\begin{lstlisting}[style=Interface]
template<class RootBasis, class... PathIndices>
auto subspaceBasis(const RootBasis& rootBasis,
                   const TypeTree::HybridTreePath<PathIndices...>& prefixPath);

template<class RootBasis, class... PathIndices>
auto subspaceBasis(const RootBasis& rootBasis, const PathIndices&... indices);
\end{lstlisting}

For example, suppose that \cpp{taylorHoodBasis} is the Taylor--Hood basis defined in~\eqref{ex:taylor-hood-basis}.
Then, the sub-tree of velocity degrees of freedom can be extracted by
\begin{lrbox}{\codebox}
\begin{lstlisting}[style=Example]
auto velocityBasis = subspaceBasis(taylorHoodBasis, _0);
\end{lstlisting}
\end{lrbox}
\begin{cppexample}\usebox{\codebox}\label{ex:velocity-basis}\end{cppexample}
and the (trivial) sub-tree of pressure degrees of freedom can be extracted by
\begin{lrbox}{\codebox}
\begin{lstlisting}[style=Example]
auto pressureBasis = subspaceBasis(taylorHoodBasis, _1);
\end{lstlisting}
\end{lrbox}
\begin{cppexample*}\usebox{\codebox}\label{ex:pressure-basis}\end{cppexample*}

\subsubsection{Multi-domain bases}\label{sec:multi-domain-basis}
Already covered by the concepts in
\Cref{sec:finite_element_trees} it is possible to describe
multi-physics setups, where the domain $\Omega$ is decomposed into
sub-domains $\Omega_i$ and each sub-domain has it's own set of basis
functions $\Lambda_i$.
The interpretation is that conceptually we define the basis on whole
$\Omega$ and implicitly assume the trivial extension of basis
functions outside $\Omega_i$ as constant $0$. This means that the
structure of the function space tree always keeps the same structure,
but in the localization it can happen that only some of the leaf nodes
actually contain local basis functions.
A first implementation of
this concept was proposed in~\cite{muething:2015} and is now
experimentally supported by \dunemodule{dune-functions}\footnote{This
  is currently under development and will be part of the 2.11 release.}.

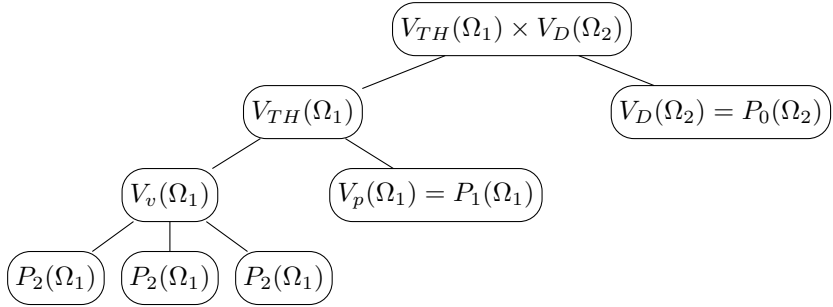
\begin{figure}
    \begin{center}
        \begin{tikzpicture}[
                level/.style={
                    sibling distance = (3-#1)*2cm + 1.5cm,
                    level distance = 1.1cm
                },
            ]
            \node [treenode] {$V_{TH}(\Omega_1) \times V_D(\Omega_2)$}
              child { node [treenode] {$V_{TH}(\Omega_1)$}
                child{ node [treenode] {$V_v(\Omega_1)$}
                    child{ node [treenode] {$P_2(\Omega_1)$} }
                    child{ node [treenode] {$P_2(\Omega_1)$} }
                    child{ node [treenode] {$P_2(\Omega_1)$} }
                }
                child{ node [treenode] {$V_p(\Omega_1) = P_1(\Omega_1)$} } }
              child{ node [treenode] {$V_D(\Omega_2) = P_0(\Omega_2)$} };
        \end{tikzpicture}
    \end{center}
    \caption{Function space tree for a multi-domain Stokes--Darcy problem.}
    \label{fig:stokes_darcy_tree}
\end{figure}

An example for such a multi-physics multi-domain problem is a coupled
Stokes--Darcy problem. A corresponding space is sketched in
\Cref{fig:stokes_darcy_tree} and setting up its basis in
\dunemodule{dune-functions} reads
\begin{lstlisting}[style=Example]
auto stokesSubDomain = @@<...>@@;
auto darcySubDomain =  @@<...>@@;
auto stokesSpace = composite(power<dim>(lagrange<2>()), lagrange<1>());
auto darcySpace = lagrange<0>();
auto stokesDarcyBasis = makeBasis(gridview,
                composite(
                  restrict(stokesSpace, stokesSubDomain),
                  restrict(darcySpace, darcySubDomain)
                )
              );
\end{lstlisting}

The original construction of a basis is extended here by the
\cpp{restrict} operation, which changes the definition in such a way,
that the Stokes sub-tree only contains local basis functions in the
\cpp{stokesSubDomain} and vice versa.

\subsection{Combining global bases and coefficient vectors}\label{sec:global-basis-and-coefficients}

Function space bases and coefficient vectors are combined to yield discrete functions by the linear combination shown exemplary in~\eqref{eq:linear_combination}.
Such discrete functions can then, for example, be written to a file or passed to a post-processing routine.
Conversely, discrete or non-discrete functions can be projected onto the span of a basis, yielding the corresponding coefficient vector.
In \dunemodule{dune-functions}, this process is called \emph{interpolation}, although it is not always interpolation in the strict mathematical sense.

To support both operations---constructing discrete functions from coefficient vectors or obtaining coefficient vectors through interpolation---it is necessary to associate individual basis functions with entries in a container that stores the coefficients.
While this mapping is straightforward in theory, it becomes technically involved in practice.
The reason lies in the nature of the multi-index types used by \dunemodule{dune-functions} to label basis functions, which are typically not directly compatible with standard random-access containers.
Furthermore, different linear algebra backends define different interfaces for accessing and modifying vector entries.

To bridge this gap, \dunemodule{dune-functions} offers several layers of abstraction that facilitate working with coefficient vectors and function space bases across various container types.
The next three sections describe these mechanisms in detail: how to access entries using vector backends (\Cref{{subsec:backends}}), how to interpret coefficient vectors as finite element functions (\Cref{subsec:discrete-global-basis-function}), and how to perform interpolation into a basis (\Cref{subsec:interpolation}).
Finally, \Cref{subsec:container-descriptors} introduces the container descriptor utility, which allows users to automatically construct coefficient containers compatible with a given basis.

Before delving into the details, we present an example of a data structure compatible with the \cpp{taylorHood} basis introduced in~\eqref{ex:taylor-hood-basis}, using a \cpp{BlockedInterleaved} index merging strategy for the velocity components and a \cpp{BlockedLexicographic} strategy for the velocity--pressure node.
A suitable container can be constructed using data structures provided by the \dunemodule{dune-istl} module, for example:
\begin{lrbox}{\codebox}
\begin{lstlisting}[style=Example]
using VectorType = MultiTypeBlockVector<
  BlockVector<FieldVector<double,dim>>, // velocity block
  BlockVector<double>>                  // pressure block
\end{lstlisting}
\end{lrbox}
\begin{cppexample}\usebox{\codebox}\label{ex:istl-vector-factory-type}\end{cppexample}
Here, the outer container behaves similarly to a \cpp{std::tuple}, but allows access to its components via \cpp{std::integral_constant} indices.
The inner blocks reflect the different nesting requirements of the two components: the velocity block requires two indices, while the pressure block requires only a single index.

\subsubsection{Vector backends}\label{subsec:backends}
Once an appropriate container type to hold the coefficient vector has been selected, it needs to be constructed and resized to the structure of the index tree.
The next challenge is to access and manipulate its entries using the multi-indices provided by the basis.
These tasks depend heavily on the specific container type and the linear algebra libraries in use.
To abstract over these differences, \dunemodule{dune-functions} defines the concept of a \emph{vector backend}.

A vector backend acts as a shim layer between the coefficient container and the basis, enabling multi-index-based access to container entries in a uniform manner and a resize utility to adapt its shape to the index tree.
Currently, backends are provided for containers from the C++ Standard Library and \dunemodule{dune-istl}.
Support for other libraries such as PETSc~\cite{petsc}, Eigen~\cite{eigen}, or Trilinos~\cite{trilinos} can be added with minimal effort.

There are two parts to the vector backend concept:
When interpreting a vector of given coefficients with respect
to a basis,  access is only required in a non-mutable way and
is defined by the \cpp{ConstVectorBackend}. It
solely requires direct access by \cpp{operator[]} using the multi-indices
provided by the function space basis:
\begin{lstlisting}[style=Interface]
auto operator[](Basis::MultiIndex) const;
\end{lstlisting}

For interpolation mutable access is required and it must be possible
to resize the vector to match the index tree generated by the basis.
These two additional methods make up the
\cpp{VectorBackend} concept:
\begin{lstlisting}[style=Interface]
auto operator[](Basis::MultiIndex);
void resize(const Basis&);
\end{lstlisting}
Note that the argument of the \cpp{resize}
member function is not a number, but the basis itself. This
is necessary because resizing nested containers requires information about the whole
index tree.

For the vector types implemented in the \dunemodule{dune-istl} and \dunemodule{dune-common}
modules, and also for some types from the C++ Standard Library, such a backend can be created using
\begin{lstlisting}[style=Interface]
template<class SomeDuneISTLVector>
auto istlVectorBackend(SomeDuneISTLVector& u);
\end{lstlisting}
where we have to assume that all leaf element types of the passed vector are identical.

While vector backends are relatively straightforward to implement, extending the concept to matrix backends requires a different approach.
In this context, combinations of dense and sparse matrices are common.
Although individual index access could, in principle, be handled using a multi-index resolution similar to that used in vector backends, this approach is often inefficient---particularly for sparse matrices.
Moreover, providing only a \cpp{resize} function is insufficient.
Instead, a suitable backend should support the definition of sparsity patterns and provide scatter-like methods for inserting multiple entries at once.
These aspects are beyond the scope of this discussion.

A reference implementation of matrix backends for data structures from Eigen, PETSc, and  \dunemodule{dune-istl} is available in the \dunemodule{dune-assembler} module~\cite{dune-assembler}, which also includes corresponding vector backends for these frameworks.

\subsubsection{Interpreting coefficient vectors as finite element functions}\label{subsec:discrete-global-basis-function}
With a container and an appropriate vector backend in place, the next step is to combine a coefficient vector and a basis into a discrete finite element function. For this purpose, \dunemodule{dune-functions} provides the utility
\begin{lstlisting}[style=Interface]
template<class Range, class B, class C>
auto makeDiscreteGlobalBasisFunction(const B& basis, const C& coefficients);
\end{lstlisting}
Given a basis \cpp{basis} and a coefficient container \cpp{coefficients}, this function returns an object that represents the corresponding finite element function.
The resulting object satisfies the \cpp{GridViewFunction} concept for the grid view associated with the basis and has the range type \cpp{Range}.
The \cpp{basis} may either be a full global basis or a \cpp{SubspaceBasis}.
In the latter case, the coefficient vector must still match the full basis structure, but only
the coefficients associated with the subspace basis functions will be used.

In general the type \cpp{C} needs to satify the
\cpp{ConstVectorBackend} concept or must be convertible to the \cpp{ISTLVectorBackend}.

Note that the range type can in general not be
determined automatically from the basis and coefficient type
because there are multiple possible types to implement this.
For example a scalar function could return \cpp{double}
or \cpp{FieldVector<double,1>}.
Hence the range type \cpp{Range} has to be given explicitly
by the user.

To give an example how \cpp{makeDiscreteGlobalBasisFunction} is used,
we again construct a \cpp{taylorHoodBasis} as in~\eqref{ex:taylor-hood-basis}.
First, we construct a subspace basis, \cpp{velocityBasis}, referring to the velocity node in the basis tree, as in~\eqref{ex:velocity-basis}.
Then, we introduce a vector of appropriate nesting structure \cpp{VectorType} associated to the full basis, as in~\eqref{ex:istl-vector-factory-type}.
Finally, we define the velocity function's range to be a \cpp{FieldVector} of size matching the number of velocity components.
All of this is combined to create a discrete function representing the velocity field in the Taylor--Hood function space:
\begin{lstlisting}[style=Example]
// Construct a container of appropriate type
auto u = VectorType{};

// Fix a range type for the velocity field
using VelocityRange = FieldVector<double,dim>;

// Create a function for the velocity field only
// but using the vector u for the full taylorHoodBasis.
auto velocityFunction
        = makeDiscreteGlobalBasisFunction<VelocityRange>(velocityBasis, u);
\end{lstlisting}
Note that the \cpp{dim} leaf nodes of the function space
tree spanned by \cpp{velocityBasis} are automatically mapped to the
\cpp{dim} components of the \cpp{VelocityRange} type.
The resulting function created in the last line implements the full \cpp{GridViewFunction}
interface described in~\cite{engwer_graeser_muething_sander:2015} and can, for example,
be directly
passed to the \cpp{VTKWriter} class of
\dunemodule{dune-grid} module to write the velocity field
as a VTK vector field, or as function parameter in the following interpolation routines.

\subsubsection{Interpolation}\label{subsec:interpolation}
The counterpart of building a discrete function from a coefficient vector and a basis is to interpolate a given function into a function space spanned by the discrete basis, i.e., to compute the coefficient vector.
Examples could be initial states in time-dependent or non-linear problems, or Dirichlet boundary values given in closed form, but need to be
transferred to a finite element representation to be usable. Depending on the finite element space,
interpolation may take different forms.  Nodal interpolation is the natural choice for Lagrange elements, but
for other spaces $L^2$-projections or Hermite-type interpolation may be more appropriate.

In \dunemodule{dune-functions} a canonical implementation is offered
by
\begin{lstlisting}[style=Interface]
template <class Basis, class C, class F>
void interpolate(const Basis& basis, C& coefficients, const F& f);
\end{lstlisting}
This method is canonical in the sense that
they use the \cpp{LocalInterpolation} functionality from
\dunemodule{dune-localfunctions} on each element for the interpolation. This is appropriate
for a lot, but not all finite element spaces. For example, no reasonable local interpolation can be defined
for B-spline  bases, and therefore the standard interpolation functionality cannot be used with the
\cpp{BSplineBasis} class.
The \cpp{interpolate} function also ensures that all necessary transformations are applied for non-affine elements. This includes, for example, the application of the chain rule for derivative degrees of freedom in Hermite-type elements, as well as Piola-type transformations for elements in $H^{\operatorname{div}}$ and $H^{\operatorname{curl}}$ spaces.

Note that this implementation requires that the range type of \cpp{f}
and the global basis \cpp{basis} are compatible. Further the type of
the coefficient vector \cpp{coefficients} either has to implement the
\cpp{VectorBackend} concept or to be wrappable by the
\cpp{istlVectorBackend}.

\dunemodule{dune-functions} implements a compatibility layer
that allows to use different vector (or matrix) types
from the \dune core modules and scalar types like, e.g., \cpp{double}
for the range of \cpp{f}, as long as the number of scalar entries
of this range type is the same as the dimension of the range space of
the function space spanned by the basis.
This also implies the assumption that the coefficients for
individual basis functions are scalar.

A variant of the
\cpp{interpolate} method is provided that allows to explicitly mark a subset of
entries to be written.
\begin{lstlisting}[style=Interface]
template <class B, class C, class F, class BV>
void interpolate(const B& basis, C& coefficients, const F& f,
                 const BV& bitVector);
\end{lstlisting}
For example, this is useful for Dirichlet values, which should only be interpolated into the constraint degrees of freedom (e.g., those associated with the Dirichlet boundary).
All other degrees of freedom must remain untouched because they might contain a suitable initial iterate obtained by other means.

The additional \cpp{bitVector} must be a container whose element type is convertible to \cpp{bool} and which mirrors the nesting structure of \cpp{coefficients}.
It indicates whether the corresponding entries in \cpp{coefficients} should be written.
For instance, in the case of flat global indices, types such as \cpp{std::vector<bool>} and \cpp{std::vector<char>} are well-suited.

As an example we consider again the \cpp{taylorHood} basis from before,~\eqref{ex:taylor-hood-basis}. We interpolate a global velocity field and a boundary velocity field in a coefficient vector, to be used for initial and boundary conditions, respectively. Let \cpp{velocityBasis} be the subspace basis associated to the velocity node,~\eqref{ex:velocity-basis}, and \cpp{u} a vector of appropriate type,~\eqref{ex:istl-vector-factory-type}.
\begin{lstlisting}[style=Example]
// A function representing a continuous velocity field
auto vel = [](const FieldVector<double,dim>& x) { return x; };

// The solution coefficient vector
auto u = VectorType{};

// Interpolate into the whole velocity subspace basis
interpolate(velocityBasis, u, vel);
\end{lstlisting}
In order to interpolate just in the boundary velocity degrees of freedom, we first need to setup the bit vector\footnote{Due to the implementation of \cpp{std::vector<bool>} using proxy types in the element access, the type \cpp{bool} cannot be used for this particular \cpp{bitVector} container in combination with the \cpp{istlVectorBacked} wrapper. The type \cpp{char} is a valid replacement in this case.}
\begin{lstlisting}[style=Example]
// A bit vector of nesting structure compatible to the basis
auto bitVector = Dune::TupleVector<
  std::vector<std::array<char,dim>>,
  std::vector<char>>{};

// Create a vector backend and prepare the bitVector
auto bitVectorBackend = istlVectorBackend(bitVector);
bitVectorBackend.resize(taylorHoodBasis);
bitVectorBackend = false;

// Fill all entries in the bitVector corresponding to velocity boundary nodes
forEachBoundaryDOF(velocityBasis, [&](auto&& index) {
  bitVectorBackend[index] = true;
});
\end{lstlisting}
The actual interpolation is then a single line:
\begin{lstlisting}[style=Example]
auto boundaryVel = [](const FieldVector<double,dim>& x) {
  return FieldVector<double,dim>{0}; };
interpolate(velocityBasis, u, boundaryVel, bitVector);
\end{lstlisting}

\subsubsection{Container descriptors}\label{subsec:container-descriptors}
While users can explicitly define container types that match the structure of a given basis, this can be cumbersome---especially for deeply nested or non-uniform index trees.
To simplify this process, the \cpp{GlobalBasis} provides a utility known as the \emph{container descriptor}.
This descriptor extracts the nesting structure of the index tree and can be used to automatically define and construct data structures that are indexable by the corresponding multi-indices.

The container descriptor is itself a (possibly nested) container, with \cpp{.size()} information in each block and an \cpp{.operator[]} to access the sub-blocks.
The size is either provided as a \cpp{static constexpr std::size_t} in case the index-tree node has a statically known degree, or as an dynamic \cpp{std::size_t} value for dynamic nodes.
Special containers are used for tree nodes where all children are identical.
Those are called ``uniform'' containers, such as \cpp{UniformArray} and \cpp{UniformVector}.
The leaf entries in the container descriptor are of type \cpp{Value}, which is a special tag acting as a placeholder for the coefficient type stored in the target container.
The container descriptor object can be retreived from a global basis \cpp{basis} using the \cpp{basis.containerDescriptor()} method or the free function \cpp{containerDescriptor(basis)}.
If the basis cannot provide an appropriate descriptor, e.g., in case of user-defined basis nodes without that description, it falls-back to the tag \cpp{Unknown}.

An actual data structure can be obtained by mapping the container descriptor blocks to specific types and the \cpp{Value} node to the coefficient type.
For data structures defined in \dunemodule{dune-istl}, this mapping is provided as a factory function:
\begin{lstlisting}[style=Interface]
template<class T = double, class CD>
auto makeISTLVector(const CD& containerDescriptor);
\end{lstlisting}
with \cpp{T} the value type of the resulting container.
It is constructed using the containers default constructor, which initializes all entries to zero.
For nested data structures build from containers of the C++ Standard Library, the factory \cpp{makeContainer} can be used similarly:
\begin{lstlisting}[style=Interface]
template<class T, class CD>
auto makeContainer(const CD& containerDescriptor, const T& defaultValue=T{});
\end{lstlisting}
Here, the value type is not defaulted to \cpp{double} and an additional overload with default value for the container entries is provided.
\footnote{The \cpp{makeISTLVector} and \cpp{makeContainer} functions will be part of the upcoming 2.11 release.}

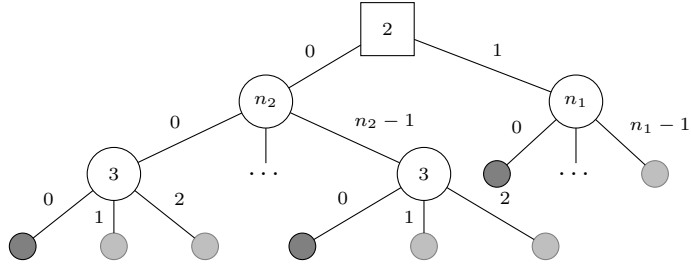
\begin{figure}
  \begin{center}
      \begin{tikzpicture}[
              level/.style={
                  sibling distance = (3-#1)*1.5cm + 1cm,
                  level distance = 1.2cm
              },
              scale=0.8,
          ]
          \node [nonuniformtreenode] {$2$}
              child{ node [uniformtreenode] {$n_2$}
                      child [sibling distance=2.5cm] { node [uniformtreenode] {$3$}
                          child [sibling distance=1.5cm] { node [leafnode] {} edge from parent node[above left,smalltext] {$0$}}
                          child [sibling distance=1.5cm] { node [inactiveleafnode] {} edge from parent node[left,smalltext] {$1$}}
                          child [sibling distance=1.5cm] { node [inactiveleafnode] {} edge from parent node[above right,smalltext] {$2$}}
                          edge from parent node[above left,smalltext] {$0$}
                      }
                      child [sibling distance=2.3cm]{ node [] {$\dots$} }
                      child [sibling distance=2.6cm]{ node [uniformtreenode] {$3$}
                          child [sibling distance=2cm] { node [leafnode] {} edge from parent node[above left,smalltext] {$0$}}
                          child [sibling distance=2cm] { node [inactiveleafnode] {} edge from parent node[left,smalltext] {$1$}}
                          child [sibling distance=2cm] { node [inactiveleafnode] {} edge from parent node[above right,smalltext] {$2$}}
                          edge from parent node[above right,smalltext] {$n_2-1$}
                      }
                      edge from parent node[above left,smalltext] {$0$}
              }
              child [sibling distance=6.2cm]{ node [uniformtreenode] {$n_1$}
                  child [sibling distance=1.3cm] { node [leafnode] {} edge from parent node[above left,smalltext] {$0$} }
                  child [sibling distance=1.3cm] { node [] {$\dots$} }
                  child [sibling distance=1.3cm] { node [inactiveleafnode] {} edge from parent node[above right,smalltext] {$n_1-1$} }
                  edge from parent node[above right,smalltext] {$1$}
              };
      \end{tikzpicture}
  \end{center}
  \caption{Container-descriptor tree for the Taylor--Hood basis with leaf blocking of velocity components.}
  \label{fig:taylor_hood_container_descriptor_blocked_tree}
\end{figure}

As an example, consider again the \cpp{taylorHoodBasis} defined in~\eqref{ex:taylor-hood-basis}. In \Cref{fig:taylor_hood_container_descriptor_blocked_tree} its container-descriptor tree is illustrated with size information in the inner nodes and filled dots for the leaf \cpp{Value} nodes. The rectangular boxes correspond to non-uniform size information, i.e., all children have a different nesting structure or shape, and the round boxes to uniform size information, i.e., all children have identical shape. For uniform nodes only a single child need to be stored.

A container of the appropriate \cpp{VectorType}, as introduced in~\eqref{ex:istl-vector-factory-type}, can be constructed directly using the factory function:
\begin{lstlisting}[style=Example]
auto u = makeISTLVector<double>(taylorHoodBasis.containerDescriptor());
\end{lstlisting}
In a similar way a \cpp{bitVector} for boundary interpolation could be provided by using a corresponding value type, e.g.,
\begin{lstlisting}[style=Example]
auto bitVector
  = makeContainer<char>(taylorHoodBasis.containerDescriptor(), false);
\end{lstlisting}
The resulting objects \cpp{u} and \cpp{bitVector} are already sized correctly to match the index tree of the basis, including all nested sub-blocks.
Therefore, no additional vector backend functionality is required to resize the containers to fit the basis.

\section{Conclusions}
We introduced concepts for composing finite element function spaces and their bases in a hierarchical manner.
This composition naturally leads to a tree structure, and exposing this structure in software interfaces offers many advantages over flat or opaque representations.
One major benefit is the ability to define flexible indexing schemes for numbering unknowns and storing associated data, such as solution vectors or stiffness matrices.
We discussed how various basic operations can be combined to construct complex indexing schemes recursively over the basis tree.

In the second part, we elaborated on how the \dunemodule{dune-functions} module implements these concepts and highlighted specific features that contribute to the flexibility and efficiency of the implementation.

We believe that the presented abstract concepts may also prove valuable for other frameworks, as they help clarify how existing implementations can be extended to support richer functionality.
Moreover, they provide a foundation for achieving interoperability between different frameworks at the level of discrete functions, represented by function space bases and their associated coefficient vectors.

\section*{Statements and declarations}


\subsection*{Ethical Approval}
not applicable.
\vspace*{-1ex}

\subsection*{Funding}
C.\ Engwer was partially supported by the German Research Foundation
(DFG) through projects HyperCut I \& II (project number 439956613),
BlockXT (project number 504505951), 
and EsCut (project number 526031774) within the DFG Priority Research Program 2410 CoScaRa ``Hyperbolic Balance Laws in Fluid Mechanics: Complexity, Scales, Randomness''. C.\ Engwer further acknowledges support by the German Research Foundation (DFG) under Germany's Excellence Strategy EXC 2044-390685587, Mathematics M\"{u}nster: Dynamics--Geometry--Structure.

S.\ Praetorius was partially supported by the German Research Foundation (DFG) through the research unit FOR3013, ``Vector- and Tensor-Valued Surface PDEs'', within the project TP06, ``Symmetry, length, and tangential constraints'' (project number 417223351).

O. Sander, C. Gräser and S. Müthing did not receive funding related to
this project.

\subsection*{Availability of data and materials}
The \dunemodule{dune-functions} gitlab
repository\footnote{\url{https://gitlab.dune-project.org/staging/dune-functions/}}
contains the code described in the implementation section. For more
details on the software stack we refere to \Cref{sec:dunefunctions}.

\subsection*{Conflict of interest}
The authors have no competing interests to declare that are relevant to the content of this article.

\bibliography{bases-concepts}

\end{document}